\documentclass{appolb}
\usepackage{epsfig}

\begin{document}
\title{The Status of the COMPASS Experiment
\thanks{Presented at DIS 2002, Cracow}%
}
\author{Mario D.~Leberig\footnote{Supported by BMBF}
\address{Institute for Nuclear Physics, Mainz University, D--55099 Mainz, Germany\\[3.ex]
  {\em on behalf of the COMPASS Collaboration}}
}
\maketitle
\begin{abstract}

One of the important goals of the COMPASS experiment is the determination of 
the gluonic contribution to the spin of the nucleon. By using the photon gluon
 fusion (PGF) process in polarised deep inelastic scattering direct access to 
the gluon polarisation can be obtained. The PGF is selected via the detection 
of open charm or high $p_T$ hadron pairs. In 2001 a large part of the 
experiment had been setup and during a two weeks period first data was taken. 
In this paper some achievements of the run in 2001 are presented together 
with first preliminary analysis results. 

\end{abstract}
\PACS{PACS numbers come here}
  
\section{Introduction}

In the recent years a series of polarised deep inelastic scattering (DIS) experiments at CERN\cite{smc}, SLAC\cite{slac} and DESY\cite{hermes} has measured the value of the singlet axial vector current matrix element $a_0$ and confirmed that the value is lower than originally expected. While in the quark parton model $a_0$ can be related to the quark contribution to the proton spin, the interpretation of $a_0$ becomes ambiguous in NLO QCD, where $a_0$ obtains contributions from the polarised gluon distribution. Generally the spin of the nucleon can be expressed in terms of the spin of the quarks $\Delta\Sigma$, of the gluons $\Delta G$ and of their orbital angular momenta L:
\begin{equation}
  \frac{1}{2} = \frac{1}{2} \Delta\Sigma + \Delta G + L_q + L_g.
\end{equation}
The measurement of the polarised gluon distribution $\Delta G$ is one 
important goal of 
the COMPASS \cite{proposal,www} experiment. In semi--inclusive DIS the photon 
gluon fusion (PGF) process allows a direct access to the gluon distribution 
in the nucleon. At COMPASS this process is singled out via the detection of 
the production of open charm or hadron pairs with high transverse momentum. 
Since the cross section for PGF is large for small $Q^2$ the full spectrum 
of virtual photons down to the quasi real $Q^2\approx 0$ region will be used. 

\section{The COMPASS Experiment}

At COMPASS we investigate deep inelastic scattering of a 160\,GeV muon beam 
on a solid state $^6$LiD or NH$_3$ target. The muon beam has an intensity of 
$2\cdot 10^8$\,muons per spill and is naturally polarised ($P\approx 80\,\%$). 
The polarised target  consists of two oppositely polarised target cells with 
a length of 60\,cm and a diameter of 3\,cm. The target cells are surrounded 
by a super conducting 2.5\,T solenoid and mounted in a cryostat to allow cooling down to 50\,mK. The polarisation is done via the dynamic nuclear polarisation method (DNP). 

\begin{figure}[htbp]
  \begin{center}
    \epsfig{file=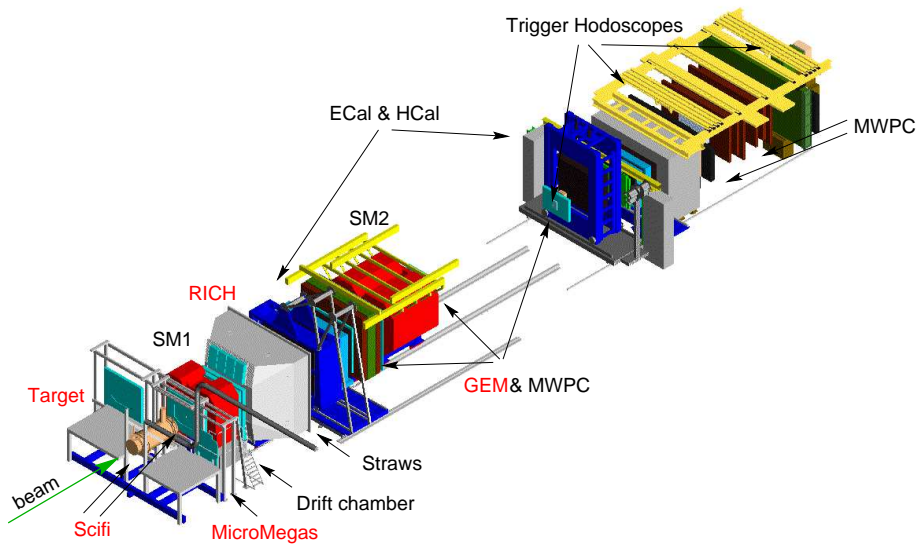,width=.96\textwidth,angle=0}
    \caption{Perspective drawing of the COMPASS spectrometer}
    \label{fig:spectrometer}
  \end{center}
\end{figure}

The COMPASS spectrometer (cf.~fig.~\ref{fig:spectrometer}) is composed out of two consecutive stages. This division is necessary to obtain a large acceptance as well as a sufficient bending power for the analysis of high energy particles. Paying tribute to the variation of particle flux in the different regions of the acceptance, a diversity of tracking detectors are used to achieve excellent resolution at reasonable costs. For small angle tracking silicon detectors, scintillating fiber hodoscopes as well as GEMs and Micromegas are used. Large angle tracking is provided by MWPCs, drift chambers (DCs) and Straw detectors. 

 For the detection of open charm production particle identification is essential. The RICH detector in the first spectrometer is designed to provide pion/kaon/proton separation for momenta between 3 and 60\,GeV/c. Additional particle identification is provided by the electro-magnetic and hadronic calorimeters as well as by two muon filters. The hadron calorimeters are further more used together with scintillator hodoscopes to build a trigger for events with quasi real photons.

Large parts of this setup had been assembled in time for the 2001 beam time. The completion of the setup has been achieved by June 2002. 

\section{The 2001 Beam Period}

In 2001 the beam period lasted for 105 days. While the first 90 days were used
 to commission the individual detectors and the central data acquisition system  the last 15 days were dedicated to a stable data taking. In the following sections the performance of various detectors will be discussed and some preliminary analysis results will be shown.

\subsection{Detector Performances} 

A major progress achieved during the setup period, was the successful polarisation of the $^6$LiD target, which was hosted in the SMC solenoid because of delivery problems of the newly designed COMPASS magnet. Polarisation values up to 58\,\% were observed for the first time in such a large target volume. In addition the spin relaxation times were measured. In the case of longitudinal target spin, where the full 2.5\,T solenoid field can be used as holding field, no relaxation could be observed during the measurement time. For the case of transversal spin orientation only a 0.5\,T holding field can be applied. Here the relaxation time was measured to be well above 1000 hours. This value is sufficiently high to enable data taking with transversely polarised target.

During the setup period we also managed to commission successfully most of the detectors. Especially the new types of tracking detectors namely scintillating fibers \cite{scifi}, GEMs \cite{gem} and Micromegas \cite{mikromegas}, were proven to work. Moreover GEMs and Micromegas achieved excellent spatial resolutions as well as very high overall efficiencies (see table \ref{tab:performance}). But also the more conventional tracking detectors like MWPCs and DCs which in 2001 collected more than 50\,\% of the tracking information showed very high efficiency as well as a good spatial resolution. 

\begin{table}[htbp]
  \begin{center}
  \begin{tabular}{|l|c|c|c|}
      \hline
                         & Scifi       & GEM        & Micromegas\\\hline
      Spatial resolution & 130\,$\mu$m & 60\,$\mu$m & $<70\,\mu$m\\
      Time resolution & 0.4\,ns     & 15\,ns     & 8\,ns\\
      Efficiency & 99\,\%      & 97\,\%     & 98\,\%\\
      Active Area  & $4\times4\,$cm$^2$ & $30\times30\,$cm$^2$ & $40\times40\,$cm$^2$\\
      \hline
    \end{tabular}
    \caption{Performance of selected tracking detectors of the COMPASS spectrometer.}
    \label{tab:performance}
    \end{center}
  \end{table}

The most delicate and complex detector in the spectrometer is the RICH1 \cite{rich}. Its radiator volume comprises 90\,m$^3$ of C$_4$F$_{10}$. The Cherenkov light is focused by 120 hexagonal mirrors onto 80000 channels of CsI cathodes which are read out via MWPCs. Since only about 50\,\% of the necessary gas volume of C$_4$F$_{10}$ could be cleaned in time for the beam time, the RICH was operated with a gas mixture of 50\% C$_4$F$_{10}$ and 50\,\% N$_2$. Despite the fact that the use of the gas mixture reduces the yield of Cherenkov photons by about a factor of two, first rings were observed. 

\subsection{Analysis Results}   

Inspired by the first data a big progress in the off-line analysis sector was achieved. First analysis results prove that the spectrometer is working as expected. 

\begin{figure}[htbp]
  \parbox[t]{.45\textwidth}{
    \epsfig{file=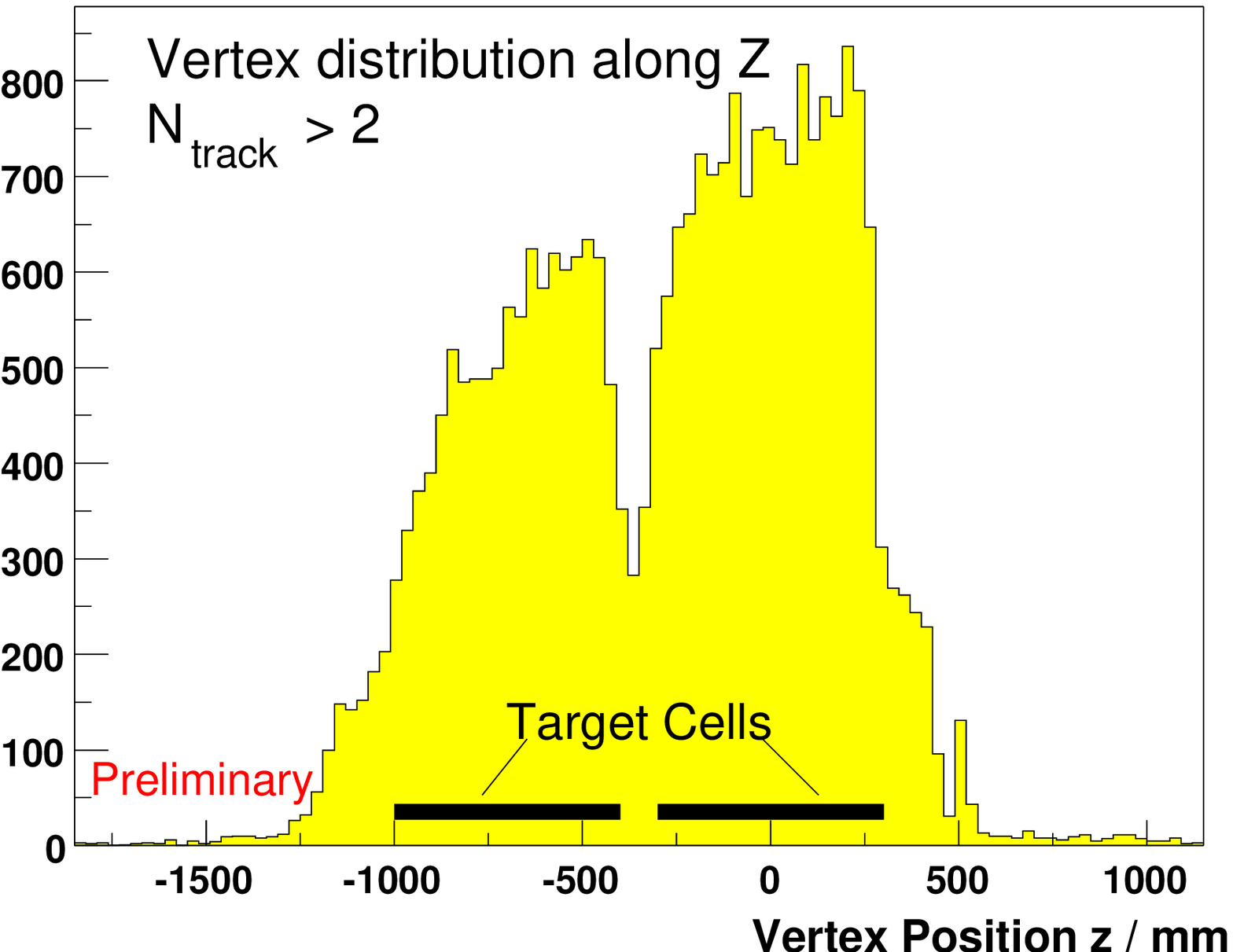,height=.35\textwidth,angle=0}
    \caption{Vertex distributions for vertices with more than one outgoing particle.}
    \label{fig:analyse1}}
  \hfill
  \parbox[t]{.45\textwidth}{
    \epsfig{file=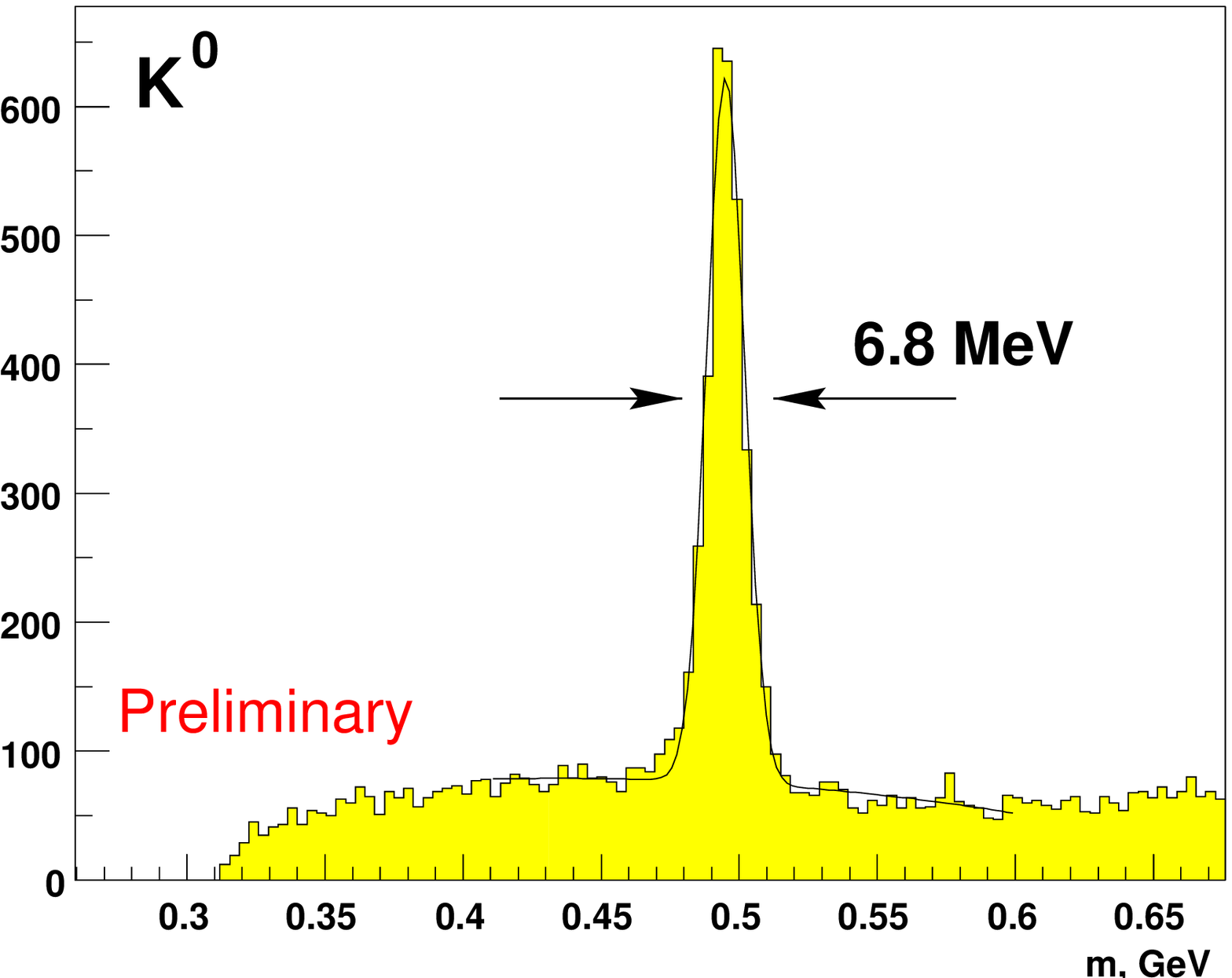,height=.35\textwidth,angle=0}
    \caption{Invariant mass spectrum for vertices lying outside the target cells.}
    \label{fig:analyse2}
    }
\end{figure}

One of the critical parameters of the COMPASS experiment is the vertex resolution along the beam direction which must be high enough to allow a clear discrimination between the two target cells separated by 10\,cm. Since the trigger accepts also events with a muon scattering angle close to zero, the detection of the scattered muon alone is not enough to accurately determine the longitudinal vertex position. However, if additional outgoing tracks are demanded, a good vertex resolution is achieved (see fig.~\ref{fig:analyse1}). For vertices with at least two outgoing tracks the most probable resolution is 1\,cm; for 85\,\% of vertices the resolution is below 5\,cm, i.e. nearly all vertices can be clearly attributed to one of the target cells. The fact that in fig.~\ref{fig:analyse1} still some vertices are found outside the target cells is explained by the presence of other material, e.g. the liquid helium and microwave stoppers, in these regions. 

Assuming that all outgoing particles (not muons) are pions an invariant mass spectrum could be produced which shows peaks at the K$^0_s$ and $\rho^0$ masses. With the hypothesis that the positive charged particle is a proton a peak at the $\Lambda$ mass is found. In fig.~\ref{fig:analyse2} the invariant mass spectrum for vertices outside the target is shown. The fit to the K$^0$ peak suggests a resolution of 6.8\,MeV which is only about 30\,\% more than predicted by simulation. 

For the run in 2002 the number of tracking detectors will be nearly doubled, hence pushing the reconstruction efficiency and the track resolution. A new trigger system and big drift chambers will be added to extend the acceptance from $Q^2=6\,$GeV$^2$ in 2001 to $Q^2=50\,$GeV$^2$ in 2002. This will allow to make simultaneous measurements with 'normal' DIS events.    

\section{Conclusion}

In 2001 large parts of the COMPASS spectrometer have been successfully commissioned. The novel detector principles used in the apparatus were proven to work. For the beam period in 2002 (about 100 days) the spectrometer will be fully operational and first physics results can be expected.

\end{document}